\documentclass[prb,aps,twocolumn,showpacs]{revtex4}
\usepackage{graphicx,color,latexsym}
\usepackage{dcolumn}
\usepackage{amsmath,amssymb,epsf,bm}
\begin{document}
\title{Long-range  effect of a Zeeman field on the electric current through the helical metal-superconductor interface in Andreev interferometer.}
\author{A.~G. Mal'shukov}
\affiliation{Institute of Spectroscopy, Russian Academy of Sciences, Troitsk, Moscow, 108840, Russia}
\begin{abstract}
It is shown that the spin-orbit and Zeeman interactions result in phase shifts of Andreev-reflected holes propagating at the surface of  a topological insulator, or in Rashba spin-orbit-coupled two dimensional normal metals, which are in a contact with an s-wave superconductor.  Due to interference of  holes reflected through different paths of Andreev interferometer the electric current through external contacts varies depending on the strength and direction of the Zeeman field. It also depends on mutual orientations of Zeeman fields in different shoulders of the interferometer.  Such a nonlocal effect is a result of the long-range coherency caused by the superconducting proximity effect. This current  has been calculated within the semiclassical theory for Green functions in the diffusive regime,  by assuming a strong disorder due to elastic scattering of electrons.
\end{abstract}
\maketitle
\section{Introduction}

Due to a combined effect of a Zeeman field and  the spin-orbit coupling (SOC) the wave functions of Cooper pairs in s-wave superconductors acquire a  phase dependent factor. This phase is responsible for the magnetoelectric effect \cite{Edelstein}, which leads to a spontaneous supercurrent in the presence of  a nonuniform static Zeeman field \cite{Malsh island, Pershoguba,Hals}, so that the spatial distribution of this current depends in a peculiar way on coordinate variations of the field. A similar phase also characterizes the electron wave function of a normal metal placed in a contact with a superconductor, if the strong enough spin-orbit and Zeeman interactions are presented in this metal. For example, it results in a spontaneous current through a  superconductor-normal metal-superconductor Josephson junction, the so called $\varphi$-junction \cite{Krive,Reinoso,Zazunov,ISHE,Liu,Yokoyama,Konschelle} which has been observed experimentally in Ref.\onlinecite{Szombati}. These physical phenomena provide important building blocks for low dissipative spintronic applications based on interaction of magnetic and superconducting systems.

It is natural to expect that in superconductor-normal metal proximity systems the phase shift, which is induced by the Zeeman field and SOC, may be observed in the Andreev reflection \cite{Andreev}, where an electron scatters from a normal metal-superconductor interface as a hole.  Interesting possibilities for studying  the phase coherent phenomena are provided by Andreev interferometers \cite{Zaitsev,Stoof,Golubov,Lambert}. These devices have several alternative paths for incident electrons and backscattered holes. In the previous studies a phase shift between interfering  scattered waves  has been provided by a magnetic flux. On the other hand, it is important to understand, if the Zeeman field can produce the phase shift that is strong enough to result in measurable effects on   the electric current through the Andreev interferometer. This problem has not been addressed  yet.

 It is clear that a strong enough SOC is needed to produce a  magnetoelectric effect which may be effective in a system of a micron size. Indeed, some two-dimensional (2D) systems have a strong intrinsic SOC \cite{Sakano,Ast,Lesne,Song}, which results in a considerable spin splitting of electron bands. In 2D systems these spin-split bands are characterized by opposite spin helicities. However, in the practically important semiclassical regime, when the Fermi energy (chemical potential) $\mu$ is larger than SOC, the magnetoelectric effect is reduced by a competition of bands with opposite helicities which cancel each other up to the terms $\sim h_F/\mu$, \cite{ISHE} where $h_F$ is the spin orbit splitting at the Fermi energy. On the other hand, this cancelation does not occur in Dirac systems, such as surface electrons in a three dimensional topological insulator (TI), because in TI only the odd number of surface helical bands cross the Fermi energy. Therefore, it is reasonable to take  a TI wire as a basic component of the device. At the same time, it will be demonstrated that the results obtained for TI may  also be extended to a conventional 2D wire having a very strong SOC  $h_F\sim \mu$.

A simple interferometer is shown at Fig.1. Due to interference of paths through the upper and lower branches of the TI wire the electric current between the normal and superconducting leads can be varied by changing magnitudes or directions of Zeeman fields in the branches. For example,  the current might be changed by flipping a magnetization direction in one of the branches. Such a nonlocal dependence of the conductance would demonstrate a long-range phase coherence created in the the TI wire by the proximity effect at low enough temperatures. The Zeeman field in TI is assumed to be directed parallel to the $x,y$ surface of the rectangular wire. It may be created by a ferromagnetic (antiferromagnetic) insulator deposited on top of TI, or by magnetic doping. Instead of fabricating TI wires, one could deposit superconducting and normal leads, as well as magnetic films on a TI flake. We will consider in detail the former setup, although qualitative results will be valid for both.

The electric current through the interferometer will be studied within the semiclassical theory for electron Green functions \cite{Eilenberger,Larkin semiclass}. A strong elastic scattering on impurities will be assumed in the TI wire, so that the corresponding mean free path is much smaller than its dimensions. Also, the elastic scattering rate is much larger than the Zeeman splitting, but much less than  the chemical potential. At the same time, for sufficiently short wires  in the micrometer range, the low-temperature inelastic scattering of electrons will be ignored.

The article is organized in the following way. In Sec.II the Usadel equation and boundary conditions for the semiclassical Green function are derived for a TI wire. In Sec.III linearized  Usadel equations are derived for the case of a weak proximity effect and the analytic expression for the current is found in the low-bias regime. A summary of the results is presented in Sec.IV.

\begin{figure}[tp]
\includegraphics[width=5cm]{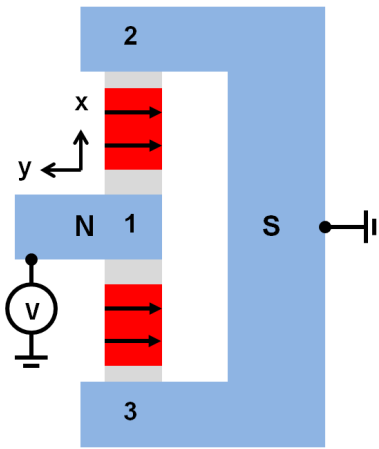}
\caption{(Color online) Andreev interferometer. "N" and "S" denote the normal and superconducting leads, respectively. The topological insulator wire contains regions having finite Zeeman fields (red) which are parallel to the TI surface (shown by arrows). These fields create a phase shift between amplitudes of Andreev reflection through the upper (12) and lower (13) branches of the TI wire. As a result, the conductance of the device oscillates as a function of this shift. It also depends on the mutual orientations of the Zeeman fields in the branches. } \label{fig1}
\end{figure}

\section{Usadel equations}

The effective one-particle Hamiltonian of  electrons on the  surface of TI can be written in the form \cite{Qi RMP}
\begin{equation}\label{H}
H=\tau_3v\mathbf{e}(\mathbf{\hat{k}}\times\bm{\sigma})-\tau_3\mu+\mathbf{Z}(\mathbf{r})\bm{\sigma}+V(\mathbf{r}),
\end{equation}
where $\mathbf{\hat{k}}=-i\partial/\partial\mathbf{r}$ and the Pauli matrices  $\tau_1,\tau_2,\tau_3$ operate in the  Nambu space, so that the electron destruction operators in the chosen basis have the form $\psi_{\uparrow},\, \psi_{\downarrow},\, \psi^{+}_{\downarrow},\,-\psi^{+}_{\uparrow}$ with the arrows denoting spin directions. The third term in Eq.(\ref{H}) represents the Zeeman interaction, where $\mathbf{Z}(\mathbf{r})$ is parallel to the $xy$ plane (the coordinate axes are shown in Fig.1), and the last term is a random impurity potential. $\mathbf{e}$ is the unit vector which is parallel to the external normal to the wire surface. It is assumed that the wire width in the $y$-direction is much larger than its thickness in the $z$-direction. Therefore, electrons spend a relatively short time on flank surfaces. For this reason these surfaces are not taken into account in Eq.(\ref{H}).

The semiclassical Eilenberger equations for electron Green functions are obtained by expanding the Dyson equation with respect to small Fermi wavelengths, in comparison with other characteristic lengths. These equations serve for calculation of the so called semiclassical Green functions. The latter are obtained from initial Green functions by integration over the particle energy at a fixed momentum direction, which is represented by the unit vector $\mathbf{n}$. These functions are  combined into the 2$\times$2 matrix $\hat{g}_{\mathbf{n}}(\mathbf{r})$ whose components are $g_{11}=g^r, g_{22}=g^a$, $g_{12}=g^K$ and $g_{21}=0$, where $g^r, g^a$ and $g^K$ are the retarded, advanced and Keldysh functions, respectively. These functions, in turn, are matrices in spin and Nambu spaces. The  procedure for the derivation of the Eilenberger equations is well described in literature \cite{Rammer,Kopnin}.  As long as all characteristic energies are much less than the Fermi energy, transitions between bands with opposite helicities can be neglected within the semiclassical approximation. In this case the spin dependence of the Green functions is locked to a momentum direction. Therefore, the initial Eilenberger equations can be projected onto the electron or hole helical bands, depending on a location of the Fermi level. The semiclassical Green function, in turn, takes the form $\hat{g}_{\mathbf{n}}=\hat{g}_{\mathbf{n}0}(1\pm \mathbf{n}\times\bm{\sigma})/2$, where at $\mu>0$ the "+" sign must be chosen and vice versa. The function $\hat{g}_{\mathbf{n}0}$ does not depend on spin and satisfies the normalization condition $\hat{g}_{\mathbf{n}0}^2=1$. For a dirty system, where the mean free path is smaller than other lengths, the Eilenberger equations can be transformed into diffusive Usadel equations \cite{Usadel} for  the matrix $\hat{g}_0(\mathbf{r})$, which is obtained from $\hat{g}_{\mathbf{n}0}(\mathbf{r})$ by averaging over $\mathbf{n}$. By this way the Usadel equation has been obtained in Ref. \cite{Zyuzin,Bobkova,Linder} for Dirac electrons and in Ref.\cite{Houzet} for a superconductor with  Rashba SOC, which is larger than the elastic scattering rate. For the TI wire this equation can be written in the form
\begin{equation}\label{Usadel}
D_{t(b)}\tilde{\bm{\nabla}}(\hat{g}_0\tilde{\bm{\nabla}}\hat{g}_0)+i[\omega\tau_3,\hat{g}_0]=0 \,,
\end{equation}
where $\tilde{\bm{\nabla}}*=\bm{\nabla}*+i[\tau_3\mathbf{F},*]$ and the gauge-field vector components are $\mathbf{F}=\mathbf{Z}(\mathbf{r})\times\mathbf{e}_z/v$. The parameters $D_t$ and $D_b$ denote electron diffusion coefficients on the top and bottom surfaces of the wire, respectively. In general these coefficients are different, because environments and surface potentials vary at these interfaces. It is interesting to note that the  Zeeman field enters Eq.(\ref{Usadel}) in the same way as the vector potential of the magnetic field. An important difference is, however, that one can not change $\mathbf{F}$ by a gauge transformation. Therefore, it is impossible to eliminate the "longitudinal" part of $\mathbf{F}$ by such a transformation. In superconductors this part results in the so called helix phase with a spatially dependent order parameter, \cite{Edelstein,Samokhin,Kaur,Agterberg,Dimitrova,Barzykin} as well as to spontaneous supercurrents around ferromagnetic islands. \cite{Malsh island, Pershoguba,Hals}

 When the wire length is much larger than its width $w$ and  $\nabla_x g_0$ is much smaller than $w^{-1}$, the Green function will tend to distribute uniformly over the wire width (in $y$-direction). If, in addition, $g_0$ is continuous on the wire flanks, it becomes constant around its perimeter.  Let us consider the case when  $\mathbf{F}$ is zero on the bottom surface. As shown in Appendix A, by averaging Eq.(\ref{Usadel}) over $y$ it can be reduced to the one-dimensional equation
\begin{eqnarray}\label{Usadel2}
&&D\tilde{\nabla}_x(\hat{g}_0\tilde{\nabla}_x\hat{g}_0)+i[\omega\tau_3,\hat{g}_0]-\nonumber \\
&&D(\gamma_x F_x^2+\gamma_y F_y^2)(\tau_3\hat{g}_0\tau_3\hat{g}_0-\hat{g}_0\tau_3\hat{g}_0\tau_3)=0\,,
\end{eqnarray}
where $\tilde{\nabla}_x*=\nabla_x* + i(D_t/2D)[\tau_3F_x,*]$, $D=(D_t+D_b)/2$,  $\gamma_x=D_tD_b/4D^2$ and $\gamma_y=D_t/2D$. It should be noted that an equation of the same form may be obtained for a Rashba 2D electron gas with large SOC, such that $h_F\sim \mu$,  by formal replacing the constants $\gamma$ and $D_t/D_b$ with parameters from Ref.[\onlinecite{Houzet}], which depend on the ratio between the Rashba constant and the Fermi velocity.

Let us consider a weak coupling of the TI  wire to the superconducting lead through tunneling barriers, which are shown in Fig.1 at contact points 2 and 3. Therefore, Eq.(\ref{Usadel2}) has to be supplemented by boundary conditions (BC) at these interfaces. For  a 2D Dirac system the usual semiclassical BC \cite{Zaitsev BC,Kupriyanov} must be modified, as shown in Ref. [\onlinecite{Zyuzin}]. The modified BC  has the  form
\begin{equation}\label{BC}
D\hat{g}_0\tilde{\nabla}_x\hat{g}_0=\Gamma_S[\hat{g}_0,\hat{g}_s],
\end{equation}
where $\hat{g}_s$ is the Green function in the superconducting lead and $\Gamma_S$ is a tunneling parameter on the interface of TI with the superconducting lead. This parameter can be written in terms of the barrier resistance $R_b=\rho_{TI}D/2\Gamma$, where $\rho_{TI}$ is the wire resistance per unit length. \cite{Kupriyanov} The Green functions and $\mathbf{F}$ in Eq.(\ref{BC}) should be taken near barriers. If the Zeeman interaction vanishes near these interfaces, then $\mathbf{F}=0$  and Eq.(\ref{BC}) coincides with a conventional expression from Ref. \cite{Kupriyanov}. Since it is assumed that the Zeeman interaction is induced by  magnetic layers on top of TI, it may vanish or not at the contacts, depending on sample preparation. It is expected that magnetization directions of the magnetic islands in the two interferometer arms may be varied independently of each other. Therefore, these islands must be separated to some extent in branching point 1.

A tunneling contact will be also assumed at the interface of the TI wire with the normal lead at point 1. At this point the Green functions of electrons in both TI branches coincide. One more BC is an evident generalization of  Eq. (\ref{BC}) that takes into account two branches which make a contact with the normal lead. We apply here the ideas of Refs. [\onlinecite{Zaitsev,Stoof}] on how to write BC in branching points. By assuming that $\mathbf{F}=0$ at  contact  point 1, this BC can be written as
\begin{equation}\label{BC2}
D\hat{g}_0\nabla_{x_{2}}\hat{g}_0 +D\hat{g}_0\nabla_{x_{3}}\hat{g}_0=-\Gamma_N[\hat{g}_0,\hat{g}_N],
\end{equation}
where $x_{2}$ and $x_{3}$ are coordinates in the branches. They are chosen so, that $x_{2}$ and $x_{3}$  are directed from contact 1 towards respective contacts 2 and 3 with the superconductor. The tunneling parameter $\Gamma_N$ may be expressed through the barrier resistance $R_{b1}=\rho_{TI}D/2\Gamma_N$, in the same way as for the TI-S contact. For the massive normal lead one may assume that its Green function is unperturbed by a contact with the TI wire. Therefore, $\hat{g}_N^{r/a}=\pm\tau_3$

\section{Andreev reflection and electric current}

We consider the case of the low temperature $T$ and small bias voltage $V$, which are much less than the superconducting gap. Therefore, the electric current between the normal and superconducting leads is determined by the Andreev reflection. This current may be expressed via the conductance $G(\omega)$, according to the well known expression \cite{Zaitsev,Volkov}
\begin{equation}\label{j}
j=\frac{1}{e}\int d\omega \left[\tanh\frac{\omega+eV}{2k_BT}-\tanh\frac{\omega-eV}{2k_BT}\right]G(\omega)\,.
\end{equation}
Let us focus on the high barrier regime, when the barrier resistance $R_b$ at TI-S interface  is much larger than the resistance of the TI wire and the barrier resistance $R_{b1}$ at the TI-N interface. In this case $G(\omega)$ is given by \cite{Zaitsev,Volkov}
\begin{equation}\label{D}
G(\omega)=\frac{1}{8R_b}(M_2+M_3)\,,
\end{equation}
where
\begin{equation}\label{M}
M_{2(3)}=\mathrm{Tr}[(g_{0}^r\tau_3-\tau_3g_{0}^a)(g_s^r\tau_3-\tau_3g_s^a)]|_{x_{2}=L_2(x_{3}=L_3)}\,.
\end{equation}
The functions $g_{0}$  are taken in TI wires near contacts 2 and 3. $L_2$ and $L_3$ are the lengths of the wires between contact 1 and contacts 2 and 3, respectively.  We assume a massive superconducting lead whose Green function is not perturbed by a proximity to TI wires. Therefore, at both contacts these functions have the form $g^r_s=g^a_s=(-i\tau_3\omega + \tau_2 \Delta)/\sqrt{\Delta^2-\omega^2}$ for $\Delta>\omega$. At high $R_b$ the Green functions in TI are weakly perturbed by the superconductor, so that  they can be represented as sums of unperturbed functions and small corrections $\delta g^{r(a)}_{j}$, namely
\begin{equation}\label{deltaf}
g^{r(a)}_0(x_{j})=\pm\tau_3+\delta g^{r(a)}_{j}\,,
\end{equation}
where $j=2,3$. The functions $\delta g^{r(a)}_{j} \ll 1$ are the anomalous Green functions which are nondiagonal in the Nambu variables. By linearizing Eq.(\ref{Usadel2}) with respect to $\delta g^{r(a)}_{j}$ it can be transformed to
\begin{eqnarray}\label{Usadel3}
&&D\left((-1)^j\nabla_{x_{j}}+2i\tau_3\tilde{F}_{xj}\right)^2\delta g^{r(a)}_{j} \pm 2 i\omega\delta g^{r(a)}_{j}-\nonumber \\
&&4D(\gamma_x F_x^2+\gamma_y F_y^2)\delta g^{r(a)}_{j}=0\,,
\end{eqnarray}
where $\tilde{F}_{xj}=F_x (x_j)D_t/(D_t+D_b)$. In turn, boundary conditions Eq.(\ref{BC}) take the linearized form
\begin{eqnarray}\label{BC3}
D\left(\nabla_{x_{j}}+2i\tau_3\tilde{F}_{xj}\right)\delta g^{r(a)}_{j}|_{x_{j}=L_j}=\Gamma_S\tau_3 [\tau_3,g_s^{r(a)}]
\end{eqnarray}
At the same time, $M_2$ and $M_3$ become
\begin{equation}\label{M2}
M_{j}=\frac{2\Delta}{\sqrt{\Delta^2-\omega^2}}\mathrm{Tr}[(\delta g^{r}_{j}+\delta g^{a}_{j})\tau_2]|_{x_{j}=L_j}\,.
\end{equation}

\begin{figure}[bp]
\includegraphics[width=6cm]{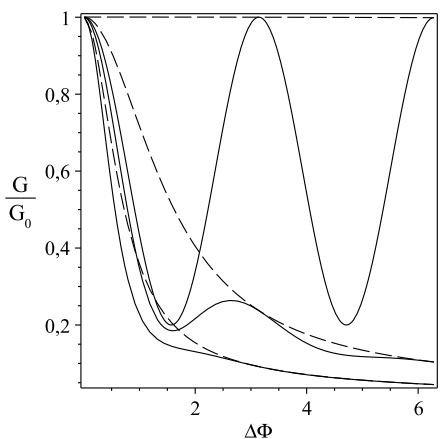}
\caption{Conductance as a function of the Zeeman field at $T=0$ and $\rho_{TI}L/R_{b1}=1$ ($G_0$ is the conductance at $Z=0$ at $T=0$), for parallel (solid) and antiparallel (dash) alignments of the Zeeman fields in TI wire branches. Curves from top to bottom : $D_b/D_t$=0, 0.1 and 0.5. } \label{fig2}
\end{figure}

The solutions of Eq.(\ref{Usadel3}) contain the phase factors $\exp(\pm 2i\int dx_{j}\tilde{F}_{xj})$ which result in spatial oscillations of Green functions. Besides these oscillations, the Zeeman interaction leads to a suppression of the superconductor proximity effect. For instance, due to the third term in Eq.(\ref{Usadel3}), $ \delta g_{2}$ and $ \delta g_{3}$ decrease with increasing  distances from contacts 2 and 3, respectively. Therefore, the length $L_Z$ of the region where $\mathbf{Z}\neq 0$ should not be too long. The corresponding condition is $2L_Z(\gamma_x F_x^2+\gamma_y F_y^2)^{1/2} \lesssim 1$. By choosing the direction of $\mathbf{Z}$ perpendicular to $x$ ($F_y=0$), the suppression effect can be reduced in samples having the  smaller ratio $D_b/D_t$ of the diffusion constants, as follows from the definition of $\gamma_x$. It is also possible to construct appropriate barriers at the flanks of the wire to guarantee a weak Klein tunneling between the top and bottom surfaces. By making the angular averaged tunneling rate much less than the Thouless energy $E_T=D/L^2$, where $L=$max$[L_2,L_3]$, the bottom surface of TI may almost completely be turned off, that will result in the small damping effect.  It should be noted that the third term in Eq.(\ref{Usadel3}) vanishes completely if the Zeeman fields are finite on both surfaces and  are equal in magnitude and antiparallel (both are perpendicular to $x$). However, such a situation is probably difficult to realize in practice.

\begin{figure}[tp]
\includegraphics[width=6cm]{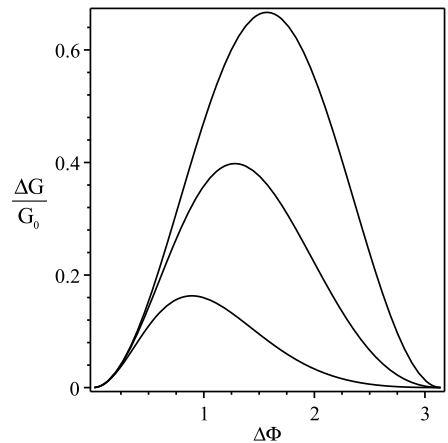}
\caption{Normalized difference of conductances for parallel and antiparallel alignments of the Zeeman fields in TI wire. From top to bottom : $D_b/D_t$=0, 0.1 and 0.5; $T=0$ and $\rho_{TI}L/R_{b1}=0.5$.  } \label{fig3}
\end{figure}

\subsection{Short wires, low bias regime}

A simple analytic result may be obtained in the case of $V\ll k_BT$  at small enough $L_2$ and $L_3$, so that $k_BT\ll E_T$. In this case one may set $\omega=0$ in $G(\omega)$ in Eq.(\ref{j}). Let us assume that  $L_2=L_3=L$ and $L_Z$ is slightly less than $L$ \cite{comment}. Hence, the phase $\Phi(x_j)\equiv2(-1)^j\int_0^{x_j} dx_{j}\tilde{F}_{xj}\simeq 2(-1)^j\tilde{F}_{xj}x_j$. The solutions of Eq.(\ref{Usadel3}) in both TI branches have the form $\delta g_{j}=\exp(i\tau_3\Phi_j)[A_j\exp(\kappa_{j} x_{j})+B_j\exp(-\kappa_{j} x_{j})]$, where $\kappa_{j}^2=4\gamma_x F_{xj}^2\pm 2i\omega$ at $F_y=0$ ($\pm$ for retarded and advanced functions, respectively). In a symmetric device, that will be assumed below for simplicity, $|F_{x2}|=|F_{x3}|$. The 2$\times$2 matrices $A$ and $B$ can be obtained from boundary conditions Eq.(\ref{BC2}), Eq.(\ref{BC3}) and the continuity of Green functions of the wire branches in contact point 1 . By substituting the so calculated $\delta g_{2}$ and $\delta g_{3}$ into Eq.(\ref{M2}) we obtain the current from Eq.(\ref{j}) in the form

\begin{equation}\label{j2}
j=\frac{\rho_{TI}}{R_b^2}\mathrm{Re}[\beta(\alpha + \cos\Delta\Phi)]V\,,
\end{equation}
where $\Delta\Phi=\Phi_2|_{x_2=L}-\Phi_3|_{x_3=L}$,
\begin{eqnarray}
&\alpha=&2(1+\Lambda)\sinh^2\kappa_0 L+1\,,  \nonumber \\
&\beta=&\frac{2}{\kappa_0(1+\Lambda)\sinh2\kappa_0 L} \,,
\end{eqnarray}
$\Lambda=\rho_{TI}\coth\kappa_0 L/2\kappa_0 R_{b1}$ and $\kappa_0=\kappa|_{\omega=0}$. For more details of the calculation, see Appendix B. It follows from these expressions that the oscillating part of the current may be of the same order as the constant term, if $\kappa L \lesssim 1$ and $\Lambda \lesssim 1$.  As can be seen from Fig.2, the current's oscillations are strongest at $D_t/D_b=0 $ and they are strongly damped already at $D_t/D_b=0.1 $. The oscillations  almost vanish at $D_t/D_b=0.5$. In the considered symmetric device the phase-dependent part of the current and the oscillations turn to zero when the Zeeman fields at two branches are antiparallel, so that in Eq.(\ref{j2}) $\Delta\Phi=0$. The difference of conductances $\Delta G$ for the parallel and antiparallel alignments is shown in Fig.3 at various ratios $D_t/D_b$ and the zero temperature ($T\ll E_T$).  An alignment switch can be performed by changing a magnetization in one of the magnetic islands. For example, one may adjust their hysteretic characteristics in such a way that an external  magnetic field of a definite strength flips the magnetization of one of them, while the other island stays in its initial state.

It is important that the considered in this subsection short wire regime is valid at low enough temperatures which provide the sufficiently large coherence length $\xi=\sqrt{D/k_BT}$, such that  $\xi \gg L$. Otherwise, one can not simply set $\omega=0$ in $G(\omega)$. Instead of that, the integral over $\omega$ in  Eq.(\ref{j}) must be taken.

\begin{figure}[tp]
\includegraphics[width=6cm]{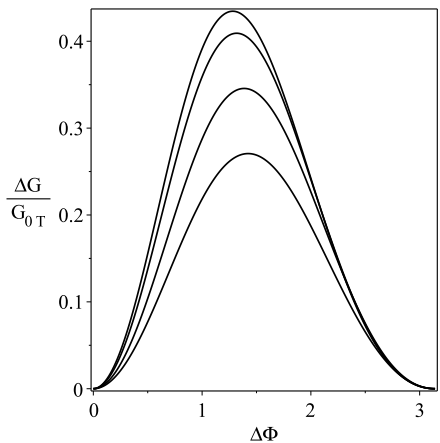}
\caption{Normalized difference of conductances for parallel and antiparallel alignments of the Zeeman fields in TI wire. $G_{0T}$ is the temperature dependent conductance of the device in the absence of the Zeeman field. Curves from top to bottom : $k_BT/E_T$ =0.1, 1, 3 and 5; $\rho_{TI}L/R_{b1}=1$ and $D_b/D_t$= 0.1} \label{fig3}
\end{figure}

\subsection{High temperatures}

In this subsection the numerical results are presented beyond the short wire regime, at $k_BT \gtrsim E_T$. $G(\omega)$ can be obtained from Eqs.(\ref{D}) and (\ref{M2}). In turn, the Green functions, that enter in Eqs.(\ref{M2}), are calculated in Appendix B. Fig.4 shows differences of conductances for parallel and antiparallel alignments of Zeeman fields, at various ratios of $k_B T$ and $E_T$. These plots are normalized by the temperature dependent conductance $G_{0T}$ in the absence of the Zeeman field. Fig.4 shows that so normalized $\Delta G$ decreases in the considered temperature interval , but not dramatically, that makes it possible to observe the phase shift produced by the Zeeman field even at relatively high temperatures. Note, that the absolute reduction of $\Delta G$ is larger, considering almost  a threefold decrease of $G_{0T}$  in the same temperature interval. In order to evaluate $E_T$, let us take the mean free path $l=10$ nm, as in Bi$_{1.5}$Sb$_{0.5}$Te$_{1.7}$Se$_{1.3}$ \cite{TIflake} and a typical Dirac velocity $v=$5 10$^5$ m/s, that gives the diffusion constant of a 2D gas $D=vl/2=$25 cm$^2/s$. With this constant the Thouless energies are   $E_T=80$mK and 20mK  for the interferometer shoulders $L=500$nm and 1000nm, respectively.

Now let us evaluate typical values of $Z$ which can provide the strong enough phase shift $\Phi =2ZLD_t/v(D_t+D_b)$. As can be seen from Figs.3 and 4, the maximum effect on $\Delta G$ is observed for $1\lesssim \Delta\Phi \lesssim 2$. For $D_t/(D_t+D_b) \simeq 1$, $v=5\cdot 10^5$m/sec and  $L= 1\mu$m the phase  $\Delta\Phi=2\Phi$ reaches 1.5 at $Z\simeq 0.1$meV. Such a field is  well below the Fermi energy, that is in agreement with the semiclassical approximation used in this work. Note, that the above evaluation of the Zeeman field is valid only for a special case of the magnetic island which covers almost the entire TI wire. Therefore, the field must be stronger for smaller sizes of the islands.

\section{Conclusion}

In conclusion, it is shown that due to a quantum interference of Andreev-scattered waves in  wires made of three dimensional TI,  the electric current through a TI-superconductor system can be varied by changing the mutual orientations of Zeeman fields in distant parts of the TI wire. This effect is a direct consequence of the long-range Cooper correlations created by the superconducting proximity effect and the Zeeman-field-induced phase shifts of the pairing functions. This effect is damped at strong Zeeman fields. The damping can be reduced by a special design of the interferometer. On the other hand, it is shown that the discussed interference effects may be observed  even at weak fields, due to the strong spin-orbit coupling of TI surface states. With some modification of parameters the theory may be extended  to ordinary two-dimensional electron systems with sufficiently strong Rashba interaction.

Acknowledgement. The work was supported by RAS Program "Actual problems of low-temperature physics".

%
%

\appendix

\section{Averaging over a wire perimeter}

Let us assume that the wire has a rectangular cross section and the coordinate $l$ runs along its perimeter, so that it coincides with $y$ and $-y$ on the top and bottom surfaces, respectively. Then, the part of Eq.(\ref{Usadel}), which is associated with the derivative over $l$, can be represented in the form:
\begin{equation}\label{nablal}
\nabla_l\left(D(l)\hat{g}_0\tilde{\nabla_l}\hat{g}_0\right)+ iD(l)\left[F_l\tau_3,\hat{g}_0 \tilde{\nabla_l}\hat{g}_0\right]               \,,
\end{equation}
where  $\tilde{\nabla_l}*=\nabla_l*+i[\tau_3F_l,*]$ and $F_l$ represents a projection of the field $\mathbf{F}$ onto the $l$ coordinate. For completeness, the lateral surfaces of the wire ($l \parallel z$) are also taken into account in Eq.(\ref{nablal}). $D(l)$ denotes the $l$-dependent diffusion constant.

 We will denote the average over the perimeter as $\overline{(...)}=\oint (...)dl/\oint dl$ and assume that $\hat{g}_0$ is constant as a function of $l$. For a diffusive transport the latter assumption is valid if the perimeter is  much smaller than the characteristic lengths which characterize variations of  Green functions  along the wire. Therefore, $\nabla_l\hat{g}_0=0$. Hence, $\tilde{\nabla_l}\hat{g}_0=i[\tau_3F_l,\hat{g}_0]$ in Eq.(\ref{nablal}). Further, since $F_l$ and $D(l)$ are periodic functions of $l$, the average of the first term in Eq.(\ref{nablal}) is 0. Therefore, the averaging of  Eq.(\ref{nablal}) yields
\begin{equation}\label{nablal2}
-\overline{D(l)F_l^2}\left[\tau_3,\hat{g}_0\left[\tau_3,\hat{g}_0\right]\right]  \,.
\end{equation}

By averaging the remaining terms in Eq.(\ref{Usadel}) over the perimeter we arrive to the one-dimensional equation
\begin{eqnarray}\label{nablal3}
\overline{D}\tilde{\nabla}_x\left(\hat{g}_0\tilde{\nabla}_x\hat{g}_0\right)+i[\omega\tau_3,\hat{g}_0]+\nonumber \\
\left(\frac{ \overline{DF_x}^2}{ \overline{D}}-\overline{DF_x^2}-\overline{DF_l^2}\right)\left[\tau_3,\hat{g}_0\left[\tau_3,\hat{g}_0\right]\right] =0       \,.
\end{eqnarray}
By assuming that the thickness of the wire is much smaller than its width one may neglect the contribution of the lateral surfaces into the average. If $F_x$ and $F_y$ are finite only on the top surface, Eq.(\ref{nablal3}) reduces to Eq.(\ref{Usadel2})

\section{Derivation of Eq.(\ref{j2})}

In each shoulder $j$ the substitution $\delta g_j=e^{i\Phi(x_j)}f_j$ , where $\Phi(x_j)\equiv2(-1)^j\int_0^{x_j} dx_{j}\tilde{F}_{xj}$,  allows to transform Eq.(\ref{Usadel3}) to the form
\begin{equation}\label{fj}
\nabla_{x_{j}}^2 f^{r(a)}_{j} +\kappa_j^2f^{r(a)}_{j}=0\,,
\end{equation}
where $\kappa^2_{j}=4\gamma_xF^2_{xj} \pm 2i(\omega/D)$ at $F_y=0$. The "$\pm$" signs in $\kappa^2$ correspond to retarded and advanced functions, respectively. Due to coordinate dependence of $F^2_{xj}$ the parameter $\kappa$ varies with $x_j$. If $F^2_{xj}$ is a step-function, the wire can be divided into several parts, so that  in each of them $\kappa^2$ is a constant. The function $f$ and its derivative must be continuous at boundaries between these parts, as it follows from Eq.(\ref{fj}). Let us consider a simple case where the homogeneous magnetic islands in each shoulder occupy almost the entire wire, except for small regions near contacts with the leads. When the lengths of these regions is much smaller than the coherence length $\sqrt{D/2|\omega|}\sim \sqrt{D/ k_BT}$, the function $f$ and its derivative are almost constant there. Therefore, by neglecting their weak spacial variation one may replace in BC (\ref{BC2}) and (\ref{BC3}) the function $f$ and $\nabla_x f$ with corresponding values in an adjacent magnetic domain. By this way it is possible to skip the small nonmagnetic regions of the wire. 

The solutions $\delta g_{2}$ and $\delta g_{3}$ of   in TI branches 12 and 13, respectively, have the form
\begin{eqnarray}\label{deltag12}
\delta g_{2}&=&e^{i\tau_3\Phi(x_2)}[A_2e^{\kappa x_{2}}+B_2e^{-\kappa x_{2}}], \nonumber \\
\delta g_{3}&=&e^{i\tau_3\Phi(x_3)}[A_3e^{\kappa x_{3}}+B_3e^{-\kappa x_{3}}]\,,
\end{eqnarray}
where the labels $r$ and $a$ are skipped for a wile. They will be restored later, where necessary. In the assumed symmetric case there is a common factor $\kappa$ in both branches. The four coefficients $A_j$ and $B_j$  anticommute with $\tau_3$ and  can be found from the boundary conditions. According to the definition of the phase $\Phi$,  we have $\Phi(0)=0$ and $\Phi(x_j)|_{x_j=L}\equiv \Phi_j$ at contacts $j=2$ and $j=3$, respectively, where $\Phi_j\simeq 2(-1)^j\tilde{F}_jL$. From the boundary conditions Eq.(\ref{BC2}), Eq.(\ref{BC3}) and the continuity of the Green functions in branches 2 and 3 at contact 1, it is easy to obtain the following equations near contact 1
\begin{eqnarray}\label{ABN}
(A_2+B_2)&-&(A_3+B_3)=0 \nonumber \\
(A_2-B_2)&+&(A_3-B_3)=\Lambda_N\delta g(0)\,,
\end{eqnarray}
where $\Lambda_N=\rho_{TI}/R_{b1}\kappa$ and $\delta g(0)=A_2+B_2=A_3+B_3$ is the Green function at contact 1.
At contacts 2 and 3 the boundary conditions have the form:
\begin{eqnarray}\label{ABS}
A_2e^{\kappa L}-B_2e^{-\kappa L}&=&e^{-i\tau_3\Phi_2}\tau_2\Lambda_S, \nonumber \\
A_3e^{\kappa L}-B_3e^{-\kappa L}&=&e^{-i\tau_3\Phi_3}\tau_2\Lambda_S\,,
\end{eqnarray}
where $\Lambda_S=(\Delta/\sqrt{\Delta^2-\omega^2})(\rho_{TI}/R_{b}\kappa)$. From equations  Eq.(\ref{ABN}), Eq.(\ref{ABS}) the factors $A$ and $B$ can be expressed as
\begin{eqnarray}\label{AB}
&&A_j=\left(\frac{\Phi_+}{\sinh\kappa L}-(-1)^j\frac{\Phi_-}{\cosh\kappa L}\right)\frac{\tau_2\Lambda_S}{4}- \nonumber \\
&&\Lambda_N\delta g(0)\frac{ e^{-\kappa L}}{4\sinh\kappa L}\,,\nonumber \\
&&B_j=\left(\frac{\Phi_+}{\sinh\kappa L}+(-1)^j\frac{\Phi_-}{\cosh\kappa L}\right)\frac{\tau_2\Lambda_S}{4}- \nonumber \\
&&\Lambda_N\delta g(0)\frac{ e^{\kappa L}}{4\sinh\kappa L}\,,
\end{eqnarray}
where $\Phi_{\pm}=\exp(-i\tau_3\Phi_3)\pm \exp(-i\tau_3\Phi_2)$.

By calculating $\delta g(0)=A_2+B_2$ from Eq.\ref{AB}  we obtain the expression for $\delta g(0)$ in the form
\begin{equation}\label{deltag}
\delta g(0)=\Phi_+ \frac{\tau_2\Lambda_S}{2\sinh\kappa L+\Lambda_N \cosh\kappa L}\,.
\end{equation}
According to Eq.(\ref{D}), the spectral conductance $G(\omega)$ is proportional to $(M_2+M_3)$. The latter may be expressed from Eq.(\ref{M2}) through the sum $\delta g^{r(a)}_2(L)+\delta g^{r(a)}_3(L)$. By substituting coefficients $A$ and $B$, that are given by Eq.(\ref{deltag}), into Eq.(\ref{deltag12}) at $x_2=x_3=L$ we obtain
\begin{eqnarray}\label{2plus3}
&&\delta g^{r}_2(L)+\delta g^{r}_3(L)+\delta g^{a}_2(L)+\delta g^{a}_3(L)=\tau_2\Lambda_S \times\nonumber \\
&&\mathrm{Re}\left[\frac{|\Phi_+|^2(\coth\kappa L+\frac{\Lambda_N}{2})}{1+\frac{\Lambda_N}{2}\coth\kappa L}+|\Phi_-|^2\tanh\kappa L\right]\,.
\end{eqnarray}
It is easy to see that in this expression only the phase difference $\Phi_2-\Phi_3$ enters, as it should be. Eq.(\ref{2plus3}) finally gives the result Eq.(\ref{j2}) in the low-bias regime where $\omega$ may be set to zero.

It is instructive to see how the phase dependence of the current vanishes in the case when only one of the two interferometer arms is conducting. Let us, for example, turn off branch 3. In this case only the second lines should be left in BC Eqs.(\ref{ABN}) and (\ref{ABS}), where $A_3=B_3=0$. It is easy to see that the solutions of these equations at $x_2=L$ have the form $A_2=\exp(-i\tau_3\Phi_2)f_a(\kappa)$ and $B_2=\exp(-i\tau_3\Phi_2)f_b(\kappa)$, where the functions $f_{a/b}(\kappa)$ do not depend on the phase $\Phi_2$. They depend only on $\kappa$. Therefore, the function $\delta g_{2}(L)$, which is given by Eq.(\ref{deltag12}), does not depend on $\Phi_2$, as well as the conductance $G$, as can be seen from Eqs.(\ref{D}) and (\ref{M2}) at $\delta g_{3}=0$.
Therefore, the only effect of the Zeeman field is a suppression of the proximity effect by the damping factor $\kappa$. It produces only a monotonous decreasing of the current at higher Zeeman fields and does not depend on its sign.


\begin{thebibliography}{99}
\bibitem{comment}
As long as characteristic lengths of $Z$ gradients near magnetic island edges, as well as distances from the island edges to contacts are much less than the  coherence length $\sqrt{D/k_BT}$ and $L$, one may represent the spacial dependence of $Z$ as a stepwise  function.
\end{thebibliography}
\end{document}